\documentclass[aps,superscriptaddress,twocolumn,twoside,floatfix,rmp,nofootinbib,a4paper,dvipsnames]{revtex4-2}
\usepackage{times}
\usepackage{amsmath}
\usepackage{amssymb}
\usepackage{amsthm}
\usepackage{bm}
\setcitestyle{square,numbers,sort&compress}
\usepackage[T1]{fontenc}
\usepackage[utf8]{inputenc}
\usepackage{dsfont}
\usepackage[caption=false]{subfig}
\usepackage{booktabs}
\usepackage[most]{tcolorbox}

\newtcbtheorem[auto counter]{ourbox}{Box}{
 colback=MidnightBlue!4!white,
 colframe=MidnightBlue!70!white,
 fonttitle=\bfseries,
 boxrule=0.4pt,
 arc=2mm,
 left=2mm,
 right=2mm,
 top=1mm,
 bottom=1mm,
 parbox=false,
 float*,
 width=\textwidth,
 label separator=.
}{obx}

\usepackage{wrapfig}

\usepackage{pifont}

\usepackage[colorlinks=true,linkcolor=blue,citecolor=magenta,urlcolor=blue]{hyperref}
\usepackage{overpic}

\theoremstyle{definition}
\newtheorem*{remark*}{Remark}

\usepackage{tabularx}
\usepackage{booktabs}

\DeclareMathOperator{\tr}{tr}

\newcommand{\ket}[1]{|#1\rangle}
\newcommand{\bra}[1]{\langle#1|}

\newcommand{\figref}[1]{Fig.~\ref{#1}} 
\newcommand{\boxref}[1]{Box~\ref{#1}} 

\usepackage[normalem]{ulem}

\begin{document}


\title{Quantum correlations in prepare-and-measure scenarios and their semi-device-independent applications}

\author{Jonatan Bohr Brask}\affiliation{Center for Macroscopic Quantum States bigQ; Department of Physics, Technical University of Denmark; Fysikvej 307; 2800 Kgs. Lyngby; Denmark}

\author{Nicolas Brunner}\affiliation{Department of Applied Physics; University of Geneva; Switzerland}

\author{Jef Pauwels}\affiliation{Department of Applied Physics; University of Geneva; Switzerland}\affiliation{Constructor University; 28759 Bremen; Germany}

\author{Davide Rusca}\affiliation{Vigo Quantum Communication Center; University of Vigo; Vigo E-36310; Spain}

\author{Armin Tavakoli}\affiliation{Physics Department and NanoLund; Lund University; Box 118; 22100 Lund; Sweden}


\begin{abstract}
A key aspect in quantum information is to understand the advantage offered by quantum systems over classical ones in communication tasks. In recent years, a fundamental approach to this problem has been developed, focusing on quantum correlations in prepare-and-measure scenarios. Inspired by the developments in Bell nonlocality and device-independent information processing, this line of research aims to characterize the possibilities and limits of quantum systems for communication, in particular to precisely capture the advantage they offer over classical systems. In addition to fundamental insights, these ideas also underpin the concept of semi-device-independent quantum information processing. Exploring trade-offs between security, performance and ease-of-implementation, this approach opens promising directions for novel quantum information processing technologies and devices. A number of protocols and proof-of-principle demonstrations have been reported in recent years, in particular for quantum randomness certification and key distribution. Here, we provide a comprehensive introduction to quantum prepare-and-measure correlations and semi-device independent applications.

\end{abstract}


\maketitle



\section{Introduction}

Quantum physics enables correlations between distant observers that cannot be explained classically. Such correlations can be established in two fundamental ways: through local measurements on pre-shared entangled states or through direct communication using quantum systems as information carriers.

Correlations arising from shared entanglement have been extensively studied in the Bell scenario, where a source distributes entangled particles between distant observers. Violations of Bell inequalities certify the presence of nonlocal correlations incompatible with classical theories respecting Bell’s notion of local causality. While Bell tests can be arranged with space-like separated observers to exclude causal influences, it is a natural question whether genuinely quantum correlations can also arise in temporally ordered experiments.
 
\begin{figure}[t!]
    \centering
    \includegraphics[width=0.9\linewidth]{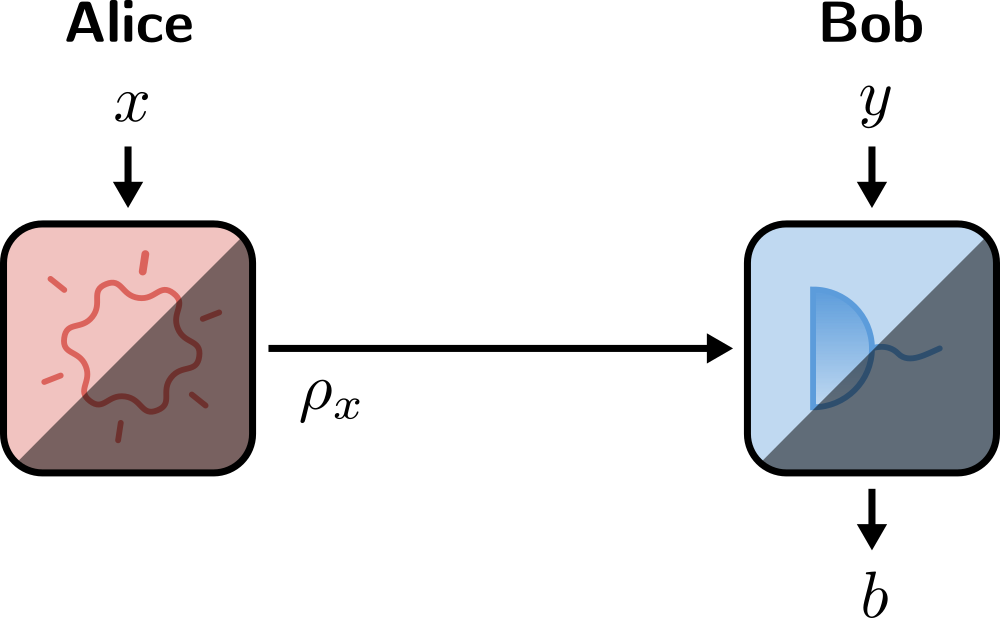}
\caption{Prepare-and-measure scenario. A preparation device (Alice) receives an input $x$ and prepares a quantum state $\rho_x$ that is sent to a measurement device (Bob) which, based on input $y$, measures the state and obtains an outcome $b$. Either the preparation or measurement (or both) may be partially characterised. In addition, the devices may have access to a shared resource which could be classical (shared randomness) or quantum (shared entangled states). The behaviour of the setup is characterised by the conditional probability distribution $p(b|x,y)$.}
\label{fig:PM_main}
\end{figure}

The prepare-and-measure scenario formalizes the most basic setting for communication and represents the simplest setup for testing quantum correlations in time. It features two separated observers, Alice the sender and Bob the receiver as shown in \figref{fig:PM_main}. Alice selects a classical symbol, $x$, and encodes it into a quantum or classical message, $\rho_x$, which is sent to Bob, who can in turn select different ‘questions’ $y$ about the incoming message, corresponding to different decoding operations. In classical systems, these correspond to data post-processing, while in quantum
systems, they involve distinct measurements $\{M_{b|y}\}_y$, with $M_{b|y} \succeq 0$ and $\sum_b M_{b|y}= \openone$, where $b$ denotes the outcome of Bob.  Quantum correlations observed in the experiment are described by a conditional probability distribution $p(b|x,y)$, given by Born's rule
\begin{equation}\label{Qcorr_noSR}
p(b|x,y)=\tr\left(\rho_x M_{b|y}\right) \,.
\end{equation}

Contrary to Bell tests, communication experiments inherently require a way to quantify and limit the amount of communication. This limitation is typically formalized as a partial assumption about the preparation device. More fundamentally, it imposes a causality constraint, analogous to the no-signalling condition in the Bell scenario, ensuring a meaningful operational separation between sender and receiver. Perhaps the simplest, and certainly the most studied approach consists in limiting the dimension, $d$, of the transmitted system. In a quantum model, this means that the set of states $\rho_x$ admits a description in the Hilbert space $\mathbb{C}^d$. In its classical counterpart, the states can be described by $d$-valued integer messages (or equivalently, a set of $d$-dimensional diagonal density matrices $\rho_x$). Different approaches limit other formal or physical aspects of the setup (such as the energy of the states $\rho_x$), motivated by practical applications and/or conceptual interest.

A central question is to understand whether and when prepare-and-measure correlations exhibit a quantum advantage, i.e., correlations of the form \eqref{Qcorr_noSR} which cannot be reproduced by a classical model. Quantum advantages in prepare-and-measure correlations enable a wide range of applications, analogous to how Bell nonlocal correlations power device-independent (DI) protocols. However, unlike DI, the need for a partial assumption about part of the setup means that a mild degree of characterisation is needed. Therefore, such protocols are called semi-device-independent (SDI).

Originally developed in the context of quantum key distribution and quantum random number generation, the SDI approach has since expanded to tasks such as quantum certification, tomography-free device characterization, and self-testing: the inference of strong constraints on the underlying implementation purely from observed correlations. SDI methods now provide a flexible framework that interpolates between fully trusted models and the black-box DI paradigm, often with substantially reduced experimental overhead.

In this article, we review these developments, providing a unified introduction to prepare-and-measure correlations and semi-device-independent methods. We begin with the fundamental structure of prepare-and-measure correlations, then discuss the various physical and information-theoretic assumptions under which quantum advantages can arise, and finally survey their main semi-device-independent applications. Our aim is to present a coherent framework that connects foundational questions about quantum correlations with practical SDI tasks.

These topics are related to many adjacent fields of research which we will not review in detail here, instead referring to existing topical reviews. This includes Bell nonlocality ~\cite{Brunner2014BellRMP}, quantum communication complexity \cite{Buhrman2010}, quantum contextuality~\cite{Budroni2022KS}, Leggett-Garg inequalities and temporal correlations  \cite{Emary2013,Schmid2024,Vitagliano2023}, device-independent quantum information processing \cite{Zapatero2023,Acin2016}, quantum random number generation\cite{HerreroCollantes2017QRNG,Ma2016QRNG,Mannalatha2023}, and quantum key distribution~\cite{Scarani2009QKD,Pirandola2020AdvancesQCrypt}.

\section{Quantum advantage}

The primary question when considering the prepare-and-measure scenario is whether quantum systems offer any advantage over classical ones. That is, are there quantum correlations of the form of \eqref{Qcorr_noSR} that cannot be reproduced classically? This question only makes sense if  we put a restriction on the system emitted by Alice, because otherwise she could simply send her input, $x$, and from it Bob could trivially generate any correlation, $p(b|x,y)$. To make a fair comparison, it is also important to consider quantum and classical systems of a similar degree of complexity. A natural approach consists in restricting to systems of a limited dimension; other approaches will be discussed in the next section. Here ``dimension'' quantifies loosely speaking the number of relevant degrees of freedom of the system, or more precisely, the number of perfectly distinguishable (i.e. orthogonal) states in which the system can be prepared. For quantum systems this corresponds to the Hilbert space dimension, while for classical systems this represents the number of distinct states, or the cardinality of the message alphabet. 

In view of this, it is natural to ask whether a quantum $d$-dimensional system (qudit) can lead to stronger correlations compared to a classical $d$-dimensional system (dit)? Quantum correlations are necessarily at least as strong as classical ones, because any classical dit strategy can be described by qudit density matrices diagonal in a shared basis. Moreover, note that this problem is typically discussed in a setting where the devices of Alice and Bob can be classically correlated, i.e., featuring so-called shared randomness, see Fig.~\ref{fig:PM_main}. It means that if two distinct sets of correlations can be achieved, then so can their classical mixture.  This is operationally relevant, since classical randomness is an easily accessible resource.

It is insightful to start exploring this question in a simple setting where Bob makes a single (fixed) measurement, i.e. Bob receives no input $y$. A simple task is for Bob to guess Alice's input $x$, that is to maximize the quantity $\sum_x p(x) p(b=x|x)$.  Ref.~\cite{Elron2007} showed that there is no quantum advantage for this task, in the sense that qudits and dits perform equally, regardless of the input prior $p(x)$, a result that can be viewed as a single-shot analogue of Holevo’s bound \cite{Holevo1973}. Later, a more general argument showed that any correlation, $p(b|x)$, generated from arbitrary qudit states and measurements, can be reproduced exactly by sending a classical dit \cite{Frenkel2015}. 

Since the above means that quantum systems cannot provide any advantage over classical ones when Bob has a fixed input, it becomes natural to explore the implications of introducing multiple inputs for Bob. In other words, Bob now also receives an input, $y$, based on which he chooses his measurements. A necessary condition for a quantum advantage becomes that the Bob's set of measurements are not jointly measurable \cite{Uola2023} since otherwise they could be reduced to post-processing of a single measurement, which we know cannot provide a quantum advantage.

\begin{figure}[t!]
    \centering
   \includegraphics[width=0.4\linewidth]{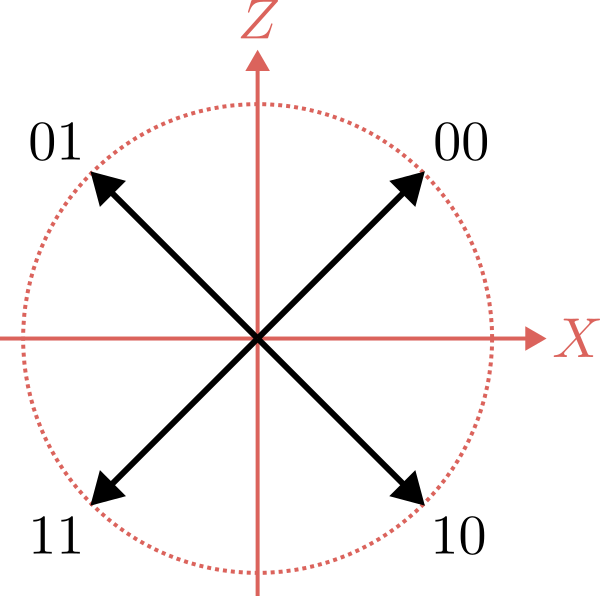}
    \caption{\emph{Optimal quantum RAC strategy}. Alice encodes her bits into qubit states forming a square rotated by \(45^\circ\) in the $XZ$ plane of the Bloch sphere. If Bob measures \(Z\) for \(y=1\) and \(X\) for \(y=2\), each term in \eqref{RAC} equals \((1 + \frac{1}{\sqrt{2}})/2\).}
    \label{fig:RAC}
\end{figure}

However, incompatible measurements for Bob open up the possibility of a quantum advantage. The paradigmatic example is the \emph{random access code} (RAC), introduced in \cite{Wiesner1983} and later rediscovered in \cite{Ambainis1999, Ambainis2002}. In its simplest form, Alice receives a (uniformly random) pair of bits \(x_1, x_2 \in \{0,1\}\), and Bob, upon receiving his input \(y \in \{1,2\}\), must guess the corresponding bit of Alice's database, i.e.~$x_y$. Performance is quantified by the average success probability: 
\begin{equation}\label{RAC}
    P_\text{RAC}=\frac{1}{8}\sum_{x_1,x_2,y}p(b = x_y \mid x_1, x_2, y) \,.
\end{equation}
Clearly, no strategy using a single binary message can achieve $P_\mathrm{RAC}=1$: for any fixed encoding into one bit (or one qubit) and any decoding rule, at least one of Bob’s two queries must be answered with error for some input pair $(x_1,x_2)$, implying $P_\mathrm{RAC}<1$. It is intuitive to see that any strategy involving a one-bit message must satisfy $P_\text{RAC} \leq \frac{3}{4}$: for example, Alice sends one of the two input bits, and Bob guesses the other bit at random. Sending a qubit message allows for higher scores. In particular, using the qubit states and measurements depicted in \figref{fig:RAC}, one can achieve 
\begin{equation} \label{Eq:qubitRAC}
P_\text{RAC} = \frac{2+\sqrt{2}}{4} \approx 0.854 > \frac{3}{4} \,
\end{equation}
which represents the highest possible score for any qubit strategy~\cite{Ambainis2009}.

This first example of a quantum-over-classical advantage in the prepare-and-measure scenario motivates many further questions. What is the minimal dimension of classical systems necessary to reproduce a given set of quantum correlations? How large can the gap between quantum and classical correlations become when considering systems of increasing dimension $d$? Do high-dimensional quantum systems lead to stronger correlations compared to low-dimensional ones?

\begin{ourbox}{The geometry of prepare-and-measure correlations}{PMgeombox}
Consider a prepare-and-measure setup as in \figref{fig:PM_main} with finite numbers $M$, $N$, $K$ of preparations, measurements, and outputs, respectively. How can we characterize all classical or quantum correlations?

To start, note that any observed correlation can be represented as an $MNK$-dimensional vector of conditional probabilities, $\bm{p} = \{p(b|x,y)\}_{b,x,y} \in \mathbb{R}^{NMK}$. The sets of classical and quantum correlations are then subsets of $\mathbb{R}^{NMK}$, that we can denote $\mathcal{C}$ and $\mathcal{Q}$, respectively, and we have $\mathcal{C} \subseteq \mathcal{Q}$. When the parties have shared randomness, these sets are convex; whenever two correlations $\bm{p}$ and $\bm{p}'$ are in $\mathcal{C}$ (or $\mathcal{Q}$) then so is any convex combination $q \bm{p} + (1-q)\bm{p}'$ with $0\leq q\leq 1$. An illustration of this geometry is shown below.

\begin{center}
    \includegraphics[width=0.75\linewidth]{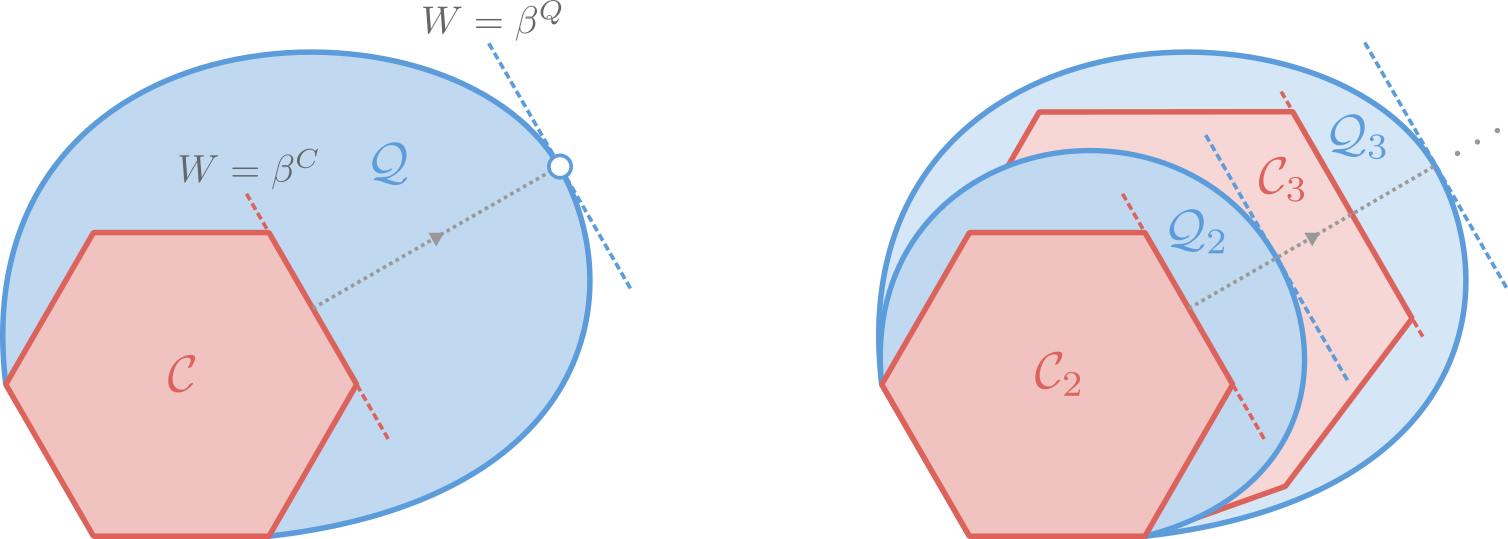}
\end{center}

\textbf{Classical set $\mathcal{C}_d$}. In a classical model, Alice encodes $x$ into a $d$-dimensional message, represented by a classical variable $m$ taking (at most) $d$ distinct values, and Bob outputs $b$, given $m$ and $y$. The resulting correlations can be decomposed
\begin{equation} \label{Eq:classicalmodel_d}
 p(b|x,y)=\sum_{\lambda} q_{\lambda}\sum_{m=1}^{d} p_A(m|x,\lambda)\,p_B(b|y,m,\lambda),
\end{equation}
where $\lambda$ denotes shared randomness (with distribution $q_\lambda\geq 0$, $\sum_\lambda q_\lambda=1$), $p_A(m|x,\lambda)$ represents the encoding of Alice, and $p_B(b|m,y,\lambda)$ the decoding of Bob. Since any local randomness can be absorbed into the shared randomness, without loss of generality, we can restrict to deterministic encodings and decodings, labelled by $\lambda$ and denoted $A_\lambda(m|x)$ and $B_\lambda(b|y,m)$, respectively. Hence, the classical set  $\mathcal C_d$ forms a convex polytope, with vertices given by deterministic correlations $D_\lambda(b|x,y) = \sum_{m=1}^d A_\lambda(m|x) B_\lambda(b|y,m)$. Membership in $\mathcal C_d$ can be efficiently checked via linear programming. Equivalently, $\mathcal C_d$ can be characterized by the facets of the polytope that provide convenient witnesses \cite{Gallego2010} (see below).

\textbf{Quantum set \(\mathcal{Q}_d\).} In the quantum case, Alice prepares \(d\)-dimensional states \(\rho_x\), and Bob performs a general quantum measurement (POVM) \(\{M_{b|y}\}\), yielding correlations of the form
\begin{equation} \label{Eq:quantum_d}
p(b|x,y) = \sum_\lambda q_\lambda\, \operatorname{tr}[\rho_x^{(\lambda)}\,M_{b|y}^{(\lambda)}],
\end{equation}
The quantum set \(\mathcal{Q}_d\) is also convex but not a polytope. However, it can still be characterized. The construction of explicit correlations provides inner approximations, whereas outer approximations can be constructed from a hierarchy of semidefinite relaxations~\cite{Navascues2015} that converge to the quantum set in many cases \cite{Navascues2015a}. Since the classical model Eq.~\eqref{Eq:classicalmodel_d} can be reproduced via a set of diagonal quantum states in a fixed basis, we have \(\mathcal{C}_d \subset \mathcal{Q}_d\), as expected.

\textbf{Witnesses.} Any convex set can be fully characterized by hyperplanes. These correspond to linear functionals of the form 
\begin{equation} \label{eq:witness}
W[\mathbf{p}] = \sum_{b,x,y} c_{b,x,y}\,p(b|x,y),
\end{equation} 
where $c_{b,x,y} \in \mathbb{R}$. The classical set (polytope) is characterized by a finite number of hyperplanes, given by inequalities of the form $W[\mathbf{p}] \leq \beta^C = \max_{\mathbf p\in\mathcal C_d} W[\mathbf p]$. The quantum set can be characterized in a similar way, except that an infinite number of hyperplanes $W[\mathbf{p}] \leq \beta^Q = \max_{\mathbf p\in\mathcal Q_d} W[\mathbf p]$ is now necessary ($\mathcal{Q}_d$ being no longer a polytope). These inequalities can be used as classical (or quantum) ``dimension witnesses'' \cite{Gallego2010}, similar to Bell inequalities. For example, for any correlation $\mathbf{p}^*$ outside of $\mathcal C_d$ (i.e. not producible with classical systems of dimension $d$) there exists a witness such that $W[\mathbf{p}^*] >\beta^C$. 

More generally, such witnesses are effective tools for certifying a quantum advantage, whenever \(\beta^{\mathrm Q} > \beta^{\mathrm C}\) for a certain functional $W$, as sketched in the left figure above. They can be interpreted as the score in a communication game, as in the RAC example of the main text (see Eq. \eqref{RAC}). 

\end{ourbox}

To address these questions, a number of methods have been developed. In particular, it is insightful to consider prepare-and-measure correlations from a geometrical perspective \cite{Gallego2010}, borrowing insights from the field of Bell nonlocality. The idea is to view each correlation $p(b|x,y)$ as a vector in a high-dimensional space. Then, the set of correlations achievable with classical (or quantum) systems of a certain dimension $d$ corresponds to a subset of this space, which can be characterized with existing techniques. The classical set is always included in the quantum set, and a quantum advantage then corresponds to a strict inclusion. These ideas are developed more formally in Box~1. 

More generally, useful connections with the scenario of Bell nonlocality can be established and exploited \cite{Brukner2004}. While the Bell scenario usually considers two parties (Alice and Bob) that are space-like separated, it can also be viewed as a prepare-and-measure setup. Here Alice acts as a preparer: by measuring half of a shared entangled state, she remotely prepares (or steers) the state of Bob. In turn, Bob performs a local measurement on the received system. An illustrative example of this connection is that the optimal strategy for the well-known Clauser-Horne-Shimony-Holt Bell inequality (involving a two-qubit Bell state and anti-commuting Pauli measurements) directly translates into the optimal qubit strategy for the RAC (discussed above), see Fig.~\ref{fig:RAC}. In fact, this mapping can be formalized for a large class of Bell inequalities such as XOR games \cite{Tavakoli2020b, Tavakoli2017z, Catani2024,baroni2025}. This link enables the transfer of many results from Bell nonlocality to the prepare-and-measure setting (and vice versa) and has been used in the area of communication complexity \cite{Buhrman2010}. Beyond XOR games, this link no longer holds in general \cite{Pawlowski2012, Tavakoli2016}.

Instances of quantum advantage have been shown for arbitrary dimensions (qudit vs dit) based on tasks of quantum state discrimination \cite{Brunner2013} and generalized random access codes \cite{Tavakoli2015rac}. In order to quantify this advantage, a common procedure investigates the noise-robustness of the quantum correlations. Specifically, one can apply a depolarisation channel to the states, $\ket{\psi_x} \rightarrow \rho_x = v \ket{\psi_x}\bra{\psi_x} + (1-v) \openone/d $, and look for the critical visibility, $v^*$, for which the quantum advantage disappears. The latter tends to increase with $d$, i.e.~the quantum advantage becomes increasingly fragile. Moreover, for a given set of states, one can identify a threshold visibility $v^*$, below which no set of measurements can generate a quantum advantage \cite{Gois2021}. A natural case is to consider the set of all possible states. It has been shown that for qubits that $v^*> \frac{1}{2}$ \cite{Bowles2015}, and for arbitrary dimensions $v^* > \frac{H_d-1}{d-1}$ where $H_d=\sum_{k=1}^{d}\frac{1}{k}$ \cite{Cobucci2025}. 

A complementary quantification of the quantum advantage is to ask for the minimal classical dimension necessary to reproduce quantum correlations, considering all possible states \cite{Cerf2000}. This problem has been solved for qubits: surprisingly, two bits of classical communication are necessary and sufficient. This was first proven for the case of projective measurements \cite{Toner2003}, and recently extended to all measurements \cite{Renner2023}. Beyond qubits, this question remains mostly open \cite{schlosser2026}. It is known that the minimal classical dimension must grow at least exponentially, which follows from results on the communication complexity of Bell nonlocal quantum correlations \cite{Brassard1999,Buhrman2010,Montina2011,Montanaro2019}. More generally, however, it is still an open question if all qudit correlations can be simulated via finite-dimensional classical systems. 
In a more general setting, featuring not only preparation and measurement devices, but also a number of intermediate transformation devices, no finite-dimensional classical systems can simulate qubit correlations \cite{Galvao2003}.

All the above questions have also been investigated in a scenario where the preparation and measurement devices are assumed to be independent, i.e.\ shared randomness is no longer available. This removes convexity (see Box~1): the classical set becomes highly nonconvex, and quantum advantages can already appear in settings where they disappear once convex mixtures are allowed \cite{Bowles2014,Vicente2017}. Such differences do not show up at the level of the linear witnesses discussed above, but they can be revealed using suitable nonlinear witnesses \cite{Bowles2014,Hayashi2006}. A useful way to formalise the resulting ``classical simulation cost'' is to view the behaviour as a (row-stochastic) communication matrix \cite{Vicente2017,Heinosaari2024}; in the absence of shared randomness, the minimal classical dimension needed to reproduce such a matrix is captured by its nonnegative rank (a quantity that is NP-hard to compute in general \cite{vavasis2007}), whereas allowing shared randomness corresponds to taking convex combinations. While several explicit examples and techniques are known \cite{Bowles2014,Vicente2017,TavakoliIndep,Batle2022,Patra2024}, no general characterisation is available.

A different form of quantum advantage has been investigated in entanglement-assisted prepare-and-measure scenarios. Here the classical shared randomness is replaced by pre-shared entanglement between Alice and Bob. By exploiting connections to Bell tests, this leads to quantum advantages already when restricting to classical communication \cite{Cleve1997,Brukner2004, Pauwels2022a}. In particular,  quantum advantages become possible without Bob having an input \cite{Frenkel2022entanglement, Vieira2023, Rout2025, Alimuddin2023}. Equivalently, one can say that the addition of entanglement reduces the communication complexity (i.e.~the amount of classical communication  required to solve certain problems) \cite{Buhrman2010}.  Recently, these ideas have been revived for quantum communication in prepare-and-measure scenarios \cite{Tavakoli2021,Pauwels2022}. Here, the addition of entanglement also boosts correlations, as illustrated via the well-known dense coding protocol \cite{Bennett1992}, where two-bits of classical communication can be perfectly transmitted by sending only one qubit. Going beyond the dense coding task, two notable new features appear. First, non-unitary encoding channels, which decohere the entanglement,  can sometimes be better than unitaries \cite{Guo2025}. Second,  the bipartite measurement performed by Bob can in some tasks  be replaced by a product measurement, thereby avoiding the complex entangled measurements seen in dense coding \cite{Piveteau2022}. This feature becomes especially significant when quantum systems have a higher-than-qubit dimension \cite{Bakhshinezhad2024,Zhang2025}. In an even more broad picture, that generalises quantum teleportation, the role of entanglement has been explored in prepare-and-measure scenarios where the inputs of Alice and the outputs of Bob are quantum states \cite{Svegborn2026}.

Finally, an interesting question is to define the notions of dimension, information capacity and communication power in a theory-independent way. In connection with the principle of information causality \cite{Pawlowski2009}, these works have identified fundamental differences between classical, quantum and postquantum theories \cite{Brunner2014a,Massar2015,dallarno2017}.

\section{Certification}

A major challenge for the adoption of quantum technologies is the certification of quantum devices, which are typically complex systems with many degrees of freedom that are difficult to model effectively. How can a user be confident that a device operates correctly, for example, prepares the desired quantum states or enables secure communication?

A standard approach to this problem involves the use of additional devices that are calibrated a priori, for example, using an external measurement device for performing quantum state tomography. However, this in turn requires the a priori calibration of these additional devices, which apparently makes the task endless. Remarkably, quantum theory offers an elegant and efficient solution to this problem. Quantum systems can be certified solely from observed correlations, without the need for characterized devices. In other words, when quantum devices are treated as black boxes, that is,  without specifying a detailed characterization of their inner workings, it is possible to certify relevant physical properties directly from the observed data. This is the essence of ``device-independent'' certification, which has been largely explored in the context of Bell nonlocality, but is also relevant for prepare-and-measure correlations.

A first property that can be characterized in the PM scenario is simply the dimension of the transmitted systems. We have seen in the previous section (see Box 1) that correlations achievable from classical systems of a certain dimension lead to limited sets of correlations, and that these sets can be effectively characterized via (linear) witnesses. These ideas can be adapted to the case of quantum systems, where the relevant dimension is the Hilbert space dimension $d$. Given observed correlations $p(b|x,y)$, what is the minimal dimension $d$ in which there exists a set of states $\rho_x$ and measurement operators $M_{b|y}$ that reproduce the correlations of the form Eq.~\eqref{Eq:quantum_d}? Note that the concept of a minimal dimension is meaningful here only when we consider a set of quantum states \footnote{Indeed, any (single) quantum state, defined on a Hilbert space $\mathcal{H}=\mathbb{C}^d$, is basically one-dimensional; it is simply a vector in $\mathcal{H}$. However, a set of states spans a subspace of $\mathcal{H}$ which has a well-defined dimension.}. 

Quantum dimension witnesses can be constructed from linear functionals of the correlations $p(b|x,y)$ \cite{Gallego2010}. An illustrative and operational class of witnesses can be constructed based on the task of quantum state discrimination. Consider that Alice prepares one of $N$ possible states $\psi_x \in \mathbb{C}^d$, and that Bob must guess which state Alice sent, with the promise that the state is either $\psi_x$ or $\psi_{x'}$. Clearly, if the dimension is large enough, $d\geq N$, the problem is trivial since $x$ can be encoded in a set of orthogonal states and hence perfectly decoded by Bob. However, when $d<N$ the task becomes more interesting and there is a limit on how well Alice and Bob can perform. Their optimal performance is captured by the family of inequalities \cite{Brunner2013}
\begin{equation}
    p_g^{N} = \sum_{x,x'=1}^N p \big(x|(x,x') \big)-p \big(x'|(x,x')\big) \leq f(N,d)\,,
\end{equation}
where we labelled the measurements $y=(x,x')$ and $f$ is a function of $N$ and $d$.
For $d>N$ these inequalities are witnesses of dimension: whenever observed correlations $p(b|x,y)$ lead to a violation of the above inequality, one can certify that the quantum systems are (at least) of dimension $d+1$. This certificate is device independent, in the sense that it is based only on observed data, without making any assumptions about the inner working of the devices. Several other classes of quantum dimension witnesses have been developed. More generally, given an arbitrary linear functional of the form in Eq. \eqref{eq:witness} in Box 1, it is possible to compute upper bounds on the largest possible value for any quantum model using $d$-dimensional systems, via a hierarchy of semidefinite programs (SDP) \cite{Navascues2015,Navascues2015a,Pauwels2022b}. 

Beyond dimension, one may also ask if more specific properties of quantum systems can be certified in the prepare-and-measure scenario. This approach has been followed for certifying sets of states that are not of stabilizer form (i.e. testing for ``magic'', a resource for quantum computation) \cite{Zamora2025}, as well as for certifying that sets of measurements are incompatible \cite{Egelhaaf2025}. More generally, one may ask whether it is possible to precisely infer what set of states and/or set of measurements are being implemented, based only on observed correlations? Remarkably, this is possible in certain cases, a property known as ``self-testing'' \cite{Mayers2004}. Self-testing has been extensively explored in the context of Bell nonlocality \cite{Supic2020}, representing a form of ``black-box tomography'', which has been adapted to the prepare-and-measure scenario \cite{Tavakoli2018}. In this case, however, we require an additional assumption on the dimension of the quantum systems; of course, without such assumption, even classical systems can reproduce any observed correlations, hence precluding any form of self-testing. 

As an illustrative example of self-testing, let us return to the random access code game, and consider that the observed data leads to the maximal qubit score, i.e. $P_{RAC}=(2+\sqrt{2})/4$ as in Eq.~\eqref{Eq:qubitRAC}. Then, under the assumption that the quantum systems are qubits, one can certify that the set of prepared states and measurements correspond  (up to a global rotation) to those represented in Fig.~\ref{fig:RAC}. This means that the optimal RAC strategy described in Fig.~\ref{fig:RAC} is in fact the unique quantum model compatible with the observed data. Such self-testing results have also been developed for measurements in mutually unbiased bases \cite{Farkas2019}. More generally, it has been shown that any set of pure states and any set of projective measurements can be self-tested in the prepare-and-measure scenario \cite{Miklin2021,Navascues2023}. As intuition suggests, a key property for a set of states or measurements to be self-tested is the fact that it is extremal. This opens the possibility of self-testing non-projective measurements, which are extremal in the space of POVMs \footnote{For a given dimension, many extremal measurements are non-projective. When the dimension is not assumed, the correlations from these measurements can be reproduced by projective ones via Naimark dilation.} \cite{Tavakoli2020c}. Self-testing results have been reported for symmetric and informationally complete (SIC) POVMs~\cite{Rosset2019,Mironowicz2019}  and more generally all extremal qubit POVMs \cite{Drotos2024}. In addition, some works have extended the prepare-and-measure scenario with an  third, intermediate, party and shown that this enables self-testing of quantum instruments \cite{Mohan2019,Miklin2020}.
 
Beyond the conceptual interest, self-testing is also relevant from a more practical perspective, for the certification of quantum devices. For this, self-testing results must be made robust to noise. Consider again implementing the RAC game with qubits, but now (more realistically) obtaining a score that is lower than the qubit maximum, i.e.~$P_{RAC}< (2+\sqrt{2})/4$. Can we still partially characterize the sets of states and measurements? Intuitively this should be possible; if the obtained score $P_{RAC}$ is close to the qubit maximum, then the states and measurements should be ``close'' to the ideal set of Fig.~\ref{fig:RAC}. This intuition is indeed correct and can be formalized by lower bounding the fidelity between the observed set of states and measurements with respect to the ideal ones, based on the observed score $P_{RAC}$ \cite{Tavakoli2018}. Robust self-testing results have also been derived for higher-dimensional systems, but typically become more sensitive to noise. Nevertheless, these methods can still be used to efficiently certify that a measurement is non-projective \cite{Tavakoli2020c, Martinez2023}, or to lower bound the output cardinality of a measurement \cite{Steinberg2021}.

\section{Quantifying Communication}

\begin{table*}[ht]
\centering
\renewcommand{\arraystretch}{1.3}
\begin{tabular}{p{4.5cm} p{5cm} p{2cm} p{5.5cm}}
\toprule
\textbf{Assumption} & \textbf{Motivation} & \textbf{Platform} & \textbf{Notes} \\
\midrule
Dimension bound \newline  & Foundational \newline (information-theoretic) & Discrete-variable & Clear information-theoretic interpretation; central in early SDI. \\ [1ex]
Almost qudit \newline\cite{Pauwels2022b} & Practical \newline(robustness to small   violations) & Universal & Corrections to theory-experiment mismatch for dimension-based SDI. \\[1ex]
Distrust framework \newline \cite{Tavakoli2021b} & Practical \newline (imperfect implementation) & Universal & Models the impact of small fidelity imperfections in the preparation used in SDI. \\[1ex]
Quantum speed limits \newline \cite{Jones2025} & Foundational (time-energy) & Universal & Randomness certification from energy-time constraints without assuming Hilbert space. \\[1ex]
Rotational symmetry \newline \cite{Jones2023} & Foundational (spatial symmetry) & Universal & Protocols certified by observed correlations and symmetry, theory-independent. \\[1ex]
Vacuum component\newline \cite{VanHimbeeck2017} & Practical \newline(measurable quantities) & Optical\newline (CV) & Bounds on observables like photon number; measurable with trusted detectors. \\[1ex]
Overlap constraints \newline \cite{Wang2019, Brask2017} & Practical \newline(coherent states), & Optical \newline(CV) & State overlaps correspond to physical distinguishability; coherent-state friendly. \\[1ex]
Information constraint\newline (guessing probability) \newline \cite{Tavakoli2022informationally, Tavakoli2020,Chaturvedi2020} & Foundational\newline (information-theoretic) & Universal & Bounds the extractable input information; implications for many SDI models. \\
\bottomrule
\end{tabular}
\caption{Summary of common semi-device-independent assumptions, including their motivation, scope of application, and key features.}
\label{tab:SDIassumptions}
\end{table*}

The characterisation of quantum correlations in the prepare-and-measure scenarios crucially relies on how we quantify and bound communication. Several approaches to this question have been proposed and explored, based on information-theoretic concepts but also on physically motivated principles. A key question driving this research is to design SDI frameworks that are well tailored to experiments, and hence directly amenable to implementations and applications. This section is devoted to these developments, while applications will be discussed in the next section.

The developments we reviewed in the last sections focused on one particular approach for bounding communication, namely the dimension-bounded model. This is indeed well motivated from both a foundational and information-theoretic perspective, bits and qubits representing basic units of classical and quantum information. Clearly, all these theoretical developments are in principle directly amenable to experiments, where correlations are inferred from observed data. A crucial point, however, consists in justifying the assumption of the physical system having a bounded dimension. A standard approach consists in adopting a finite-dimensional description of these systems. For example, in optics experiments, it is customary to represent a polarised single photon as a qubit. Such an approach has been followed in various works, demonstrating a quantum advantage in the RAC game \cite{Spekkens2009} or in sequential tests \cite{Anwer2020, Foletto2020}. 

To go beyond qubits, other degrees of freedom must be considered, such as path and orbital angular momentum. These have been used for demonstrating concepts such as dimension witnesses \cite{Hendrych2012,Ahrens2012}, self-testing protocols for states and measurements \cite{Smania2020}, including the certification of quantum instruments \cite{Tavakoli2020,Anwer2020} and non-projective measurements \cite{Martinez2023}.

While these experimental works provide elegant proof-of-principle demonstrations, they also illustrate the inherent limits of the dimension-bounded model in practice. Indeed, physical systems typically feature many degrees of freedom, and their description in terms of a finite-dimensional system raises a number of issues. This is particularly relevant in the semi-device-independent approach where one would like to use only minimal assumptions about the setup. Consider again the example of an experiment representing polarized single photons as qubits. Clearly, photons have many other degrees of freedom, which may also contain additional (and unwanted) information about the encoded input. Moreover, any realistic single-photon source inevitably produces vacuum and/or multi-photon events with some probability. All these unavoidable imperfections make it delicate to use an assumption of bounded dimension in practice. 

These issues become particularly crucial when considering possible applications of prepare-and-measure correlations. Notably, in SDI cryptographic applications (see next section) the security of the protocol will in general heavily rely on the validity of the assumption of bounded dimension, and even small deviations can have strong consequences. This question was explored in the ``almost-qudit'' framework \cite{Pauwels2022b}, which considers that the prepared states lie predominantly within a $d$-dimensional subspace, but allow for a small fraction of support outside it. Crucially, even slight deviations can drastically impact the analysis and jeopardize security. In a similar spirit, it was also shown that when the prepared states deviate slightly from the optimal ones---potentially expanding the Hilbert space dimension---correlations experiments are strongly impacted \cite{Tavakoli2021b}, leading e.g.~to significant overestimations of randomness certification.

This illustrates a key issue for semi-device-independent protocols, namely that the assumption that is used for bounding the communication should be directly verifiable experimentally. The dimension corresponds to the number of degrees of freedom available for encoding information, not something that can be measured. This motivates exploring alternative ways of bounding the communication, using physical quantities that are directly measurable in the lab. Different approaches have been proposed (see Table~\ref{tab:SDIassumptions} for an overview), mainly motivated from experimental perspectives.

A first approach consists in assuming an upper bound on the mean energy of the emitted states \cite{VanHimbeeck2017}. In practice this is appealing, since energy can in principle be directly measured using a calibrated detector, making the assumption experimentally verifiable. For example, in an optical experiment one can use a powermeter to ensure a lower bound on the average vacuum component of given optical mode. Interestingly this model also exhibits a quantum advantage, in the sense that quantum correlations are proven to be stronger than their classical counterpart. In particular, this advantage can be witnessed in the simplest possible scenario, with only two preparations and a fixed binary measurement \cite{VanHimbeeck2017} (See \boxref{obx.qrngbox}). The latter can be readily implemented in a standard quantum optics setup with coherent states, and allows for application in randomness generation (see next section). 
More generally, this approach can be extended to constraints on the expectation value of one (or more) physical observables, for example constraining the full photon-number statistics and to more inputs and outputs \cite{Carceller2024}.

A second, related approach consists in assuming a bound on the distinguishability between the emitted quantum states \cite{Brask2017}. Formally, this can be enforced by constraining the pairwise overlaps between the states. At first sight more abstract, this approach is in fact also well adapted to experiments, notably in continuous-variable optical setups, where overlaps between coherent states have a direct operational meaning. It is also closely related to the energy-bound model: when one of the prepared states is the vacuum, the overlap constraint captures the vacuum component. The overlap-based model also enables the certification of quantum randomness in the simplest possible prepare-and-measure scenario with two preparations and a single binary measurement \cite{Brask2017}. General methods have been developed, based on hierarchies of semi-definite programs, for bounding quantum correlations in the overlap-based model \cite{Wang2019}. 

These developments provide an attractive novel path for semi-device-independent protocols and implementations. In particular, compared to the dimension-bounded model, they provide two main advantages. First, they are based on assumptions that are much better tailored to experimental platforms (in particular in optics), and hence easier to verify in practice. Moreover, they reveal quantum effects (such as certified randomness generation) in a the simplest possible setting involving only two preparations and a fixed measurement. This is in stark contrast to the dimension-bounded model, which required multiple incompatible quantum measurements to witness non-classicality.

More recently, a number of works proposed to ground SDI assumptions in fundamental spacetime symmetries, such as time and rotational invariance. One leverages quantum speed limits, showing that randomness can be certified under a bound on energy uncertainty and control over the preparation time, without assumptions on the Hilbert space or Hamiltonian \cite{Jones2025,Konstantinos2025}. The other derives constraints from rotational symmetry, introducing SDI protocols certified solely by the observed correlations and a bound on an operational notion of rotation \cite{Jones2026}. These frameworks highlight how fundamental physical symmetries can underpin security in SDI scenarios.

Finally, in contrast to the above assumptions, which constrain physical properties of the preparation device or the structure of the underlying Hilbert space, the information approach \cite{Tavakoli2022informationally, Tavakoli2020,Chaturvedi2020} imposes a purely information-theoretic constraint. It directly bounds how much information about the input can, in principle, be extracted from the preparations. Operationally, this is captured by the guessing probability, which corresponds to the optimal success probability in a state discrimination task. More generally, the information approach provides a common conceptual framework for investigating all the different SDI models discussed in this section. This allows for comparing different models and for importing results from one model to another \cite{Pauwels2025}. 

\section{Random number generation and key distribution}

Quantum systems are well suited for cryptographic tasks, combining intrinsic measurement randomness with the impossibility of perfect cloning. A key aspect in this context is the level of trust placed in the devices used to manipulate quantum states. Different paradigms have been explored, from the device-dependent setting---where cryptographic devices are fully characterized---to the device-independent scenario where devices are treated as black-boxes. The SDI approach explores the intermediate regimes, where security is established based on a few mild assumptions. 

The most mature application is quantum random number generation (QRNG), arguably the first widely commercialised quantum technology. In the device-dependent setting, QRNGs are easy to implement and can reach very high rates (up to $100\,\mathrm{Gbit/s}$ has been demonstrated \cite{Bruynsteen2023}). At the other extreme, device-independent QRNG aims to certify randomness under essentially minimal assumptions \cite{Pironio2010}, and is attractive in contexts where publicly auditable, high-assurance randomness is required (e.g.\ for public randomness beacons \cite{Kavuri2025}). It requires distribution of entangled states and detection-loophole free Bell-inequality violation, which places stringent requirements on sources and detectors, limiting speed and practicality (bit rates on the order of 100 bits/s have been reported \cite{Liu2018}). SDI protocols aim to strike a balance between the ease of implementation (and hence typically also speed) of device-dependent  protocols and the ultra-high security offered by the DI approach. A number of prepare-and-measure SDI QRNG protocols have been developed which explore different combinations of assumptions on the source or measurement devices and are amenable to implementations with different techniques.

\begin{figure*}[t!]
\centering
\includegraphics[width=\linewidth]{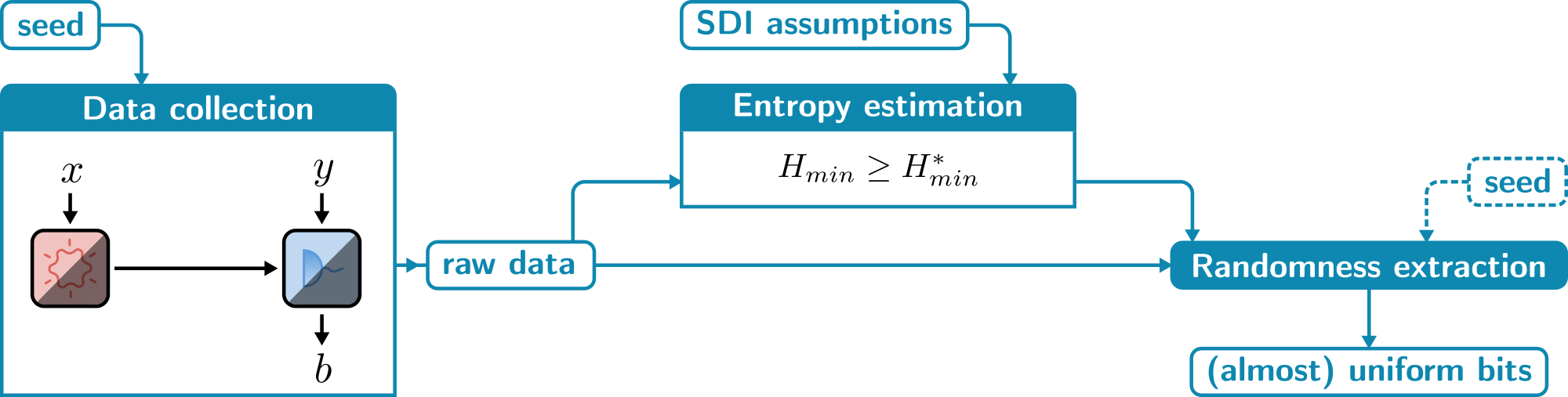}
\caption{Elements of a prepare-and-measure SDI-QRNG protocol. Data is first collected from a physical setup. The entropy in the raw data block is lower bounded based on SDI assumptions about the setup and potentially tests on the data. Finally, randomness extraction is performed, compressing the raw data to a shorter output bit string $\epsilon$-close to uniform. The data collection consumes seed randomness for choosing the preparations and measurement settings. Commonly used extractors, e.g.\ based on two-universal hashing, also require seeds.}
    \label{fig:QRNG}
\end{figure*}

\begin{ourbox}{Vacuum component SDI QRNG}{qrngbox}
A protocol for SDI QRNG based on a bound on the vacuum component can be realised e.g.\ using weak coherent states and homodyne detection. An fibre-based implementation is illustrated below (left). Alice encodes an input bit $x\in\{0,1\}$ in the phase of a weak coherent state prepared by attenuating a laser. Bob mixes the signal with a local oscillator and performs homodyne detection. The measurement outcome $q$ is binarized by sign binning to produce the output bit $b\in\{0,1\}$. This setup is simple and requires only standard optical components with no need for photon-number-resolving detectors. In the low-energy regime, the prepared states necessarily have significant overlap and cannot be perfectly discriminated, which enables randomness certification from the observed correlations. A similar scheme reported in Ref.~\cite{Rusca2019} (using displaced single-photon detection) achieved a secret bit rate of $1.25\,\mathrm{Mbit\,s^{-1}}$.
\begin{center}
    \includegraphics[width=0.98\linewidth]{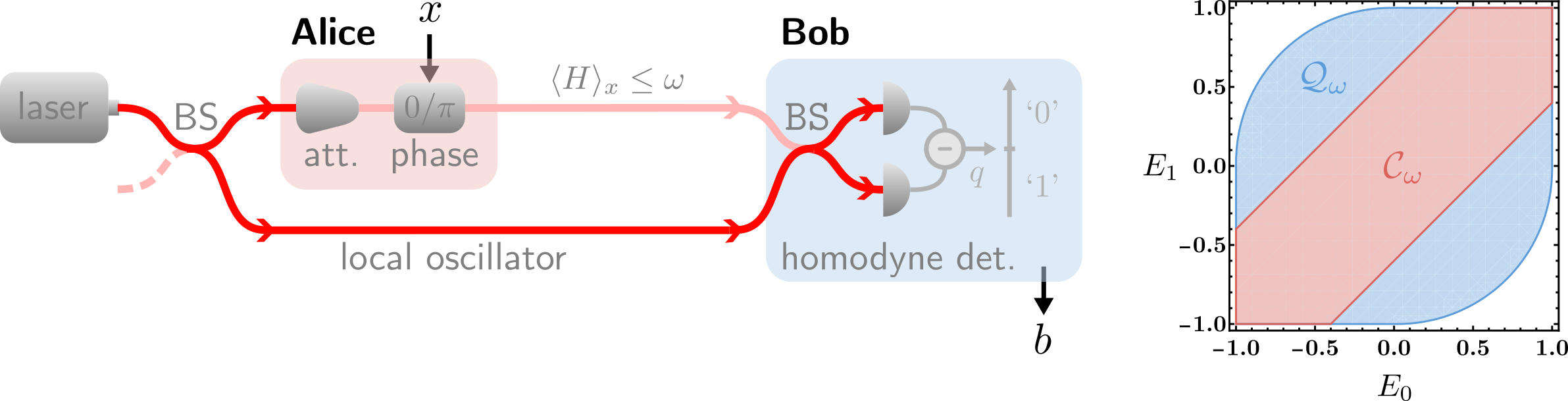}
\end{center}
In this protocol, the SDI assumption is a lower bound on the vacuum component of the transmitted system. This is operationally well motivated and directly verifiable using, e.g. a trusted powermeter. Assuming the same upper bound $\omega$ on the energy for all preparations, for any measurement $\{M_b\}$ the set $\mathcal Q_{\omega}$ of quantum correlations is given by
\begin{equation}
  p(b|x)=\tr(\rho_x M_b),\qquad \textrm{subject to} \qquad \langle H\rangle_x = \tr(H\rho_x)\le \omega \quad \forall x ,
  \label{eq:box2-core}
\end{equation}
where $\rho_x$ are arbitrary states and $\langle H\rangle_x = \tr(H\rho_x)$ the corresponding mean energies. $H$ represents the Hamiltonian of the transmitted system, which is assumed to be gapped. For binary inputs, the states can be described as qubits without loss of generality and we can take $H=\openone-\ket{0}\bra{0}$, which is the complement of the vacuum component.

\textbf{Quantum advantage.} Energy-bounded quantum correlations are strictly stronger than classical ones. This can be seen following a geometrical approach as in \boxref{obx.PMgeombox}. For a given energy bound $\omega$, the classical set $\mathcal C_{\bm \omega}$ consists of correlations achievable when the preparations are all diagonal in a common basis: $\rho_x=\sum_k q_x(k)|k\rangle\langle k|$. As there are only finitely many deterministic behaviours, $\mathcal C_{\bm \omega}$ is a polytope. In particular, for binary inputs and outputs, writing $E_x\equiv p(1|x)-p(0|x)$, the facets of this polytope are given by
\begin{equation}
  |E_0-E_1|\,\le\,4\omega\quad\Longrightarrow\quad W \leq W_{\mathrm{cl}}=\tfrac{1}{2}+\omega,
  \label{eq:box2-classical-bound-final}
\end{equation}
where the witness $W\equiv\tfrac{1}{2}\big[p(0|0)+p(1|1)\big]$ represents the average success probability for guessing $x$ from $b$.

The classical and quantum sets are depicted above (right) for $\omega=0.15$. As can be seen, the quantum set is a strict superset of the classical one, i.e $ \mathcal C_{\omega} \subsetneq \mathcal Q_{\omega}$, implying a quantum advantage. In particular, quantum correlations can violate the witness $W$:
\begin{equation}
  W \leq W_{\mathrm{cl}} < W_{\mathrm{q}}=\tfrac{1}{2}+\tfrac{1}{2}\sqrt{\omega(1-\omega)}\,.
  \label{eq:box2-quantum-bound-final}
\end{equation}
where $W_{\mathrm{q}}$ is a tight bound on the quantum value.
More generally, the classical and quantum sets can be characterized via SDP relaxations, for example in scenarios with more inputs, outputs, or additional photon-number constraints \cite{Carceller2024}.

\textbf{Randomness certification.} The SDI QRNG protocol uses the witness value $W$ as a security parameter. Observed $W> W_{\mathrm{cl}}$ implies a quantum advantage and thus the possibility to extract randomness. In practice, from the observed (finite) data, one first estimates a lower bound $W\ge W^\ast$. One then bounds the probability with which an adversary can guess Bob’s output, over all realizations compatible with the vacuum-constrained model and the observed witness value. In the simplest approach, assuming i.i.d., the problem is given by
\begin{equation}
p_g^\star=\max_{\{\rho_x,M_b\}} p_g
\quad\text{s.t.}\quad p(b|x)\in\mathcal Q_{\omega} \quad \textrm{and} \quad
 W\ge W^\ast ,
\label{eq:box2-pgstar-final}
\end{equation}
The certified single-round randomness is then quantified by the min-entropy $h_{\min}=-\log_2 p_g^\star$.  In the present binary scenario, the optimization can be solved analytically \cite{VanHimbeeck2017}. More generally, one aims to bound the block min-entropy $ H_{min}(\boldsymbol{X}|\boldsymbol{E})$ (see main text) without i.i.d. assumptions, which can be done via SDP relaxations \cite{Carceller2024}.

\end{ourbox}

The general structure of a prepare-and-measure QRNG protocol is shown in \figref{fig:QRNG}. It consists of three components: i) a physical prepare-and-measure setup generating the raw data, ii) a method for bounding the entropy in the raw data, and iii) randomness extraction via postprocessing, compressing the raw data to a shorter string of (almost) perfectly uniform bits. The security of the protocol depends on the assumptions used in the derivation of the entropy bound and on how well-justified those assumptions are for the concrete physical implementation. The speed depends on the physical repetition rate, the entropy, and the complexity of randomness extraction (for real-time extraction), which can be a bottleneck in practice when large block-sizes of data need to be processed.

A central quantity in QRNG is the (conditional) min-entropy of the raw data relative to a potential adversary. For a classical variable $X$ with distribution $p_X$, it is defined as
\begin{equation}
H_{\min}(X)=-\log_2 p_g,\qquad p_g:=\max_x p_X(x),
\end{equation}
where $p_g$ is the optimal guessing probability. In a QRNG protocol one instead considers a string of outcomes $\mathbf{X}$ and adversarial side information $\mathbf{E}$, and aims to lower bound $H_{\min}(\mathbf{X}|\mathbf{E})$ from the observed data and the chosen SDI assumptions. This quantity directly controls how many nearly uniform bits can be distilled from the raw output: privacy amplification (implemented via standard strong seeded extractors, e.g.\ two-universal hashing) compresses $\mathbf{X}$ to a shorter string that is $\varepsilon$-close to uniform and essentially independent of $\mathbf{E}$ \cite{Renner2004,Tomamichel2011,Tomamichel2016}. Establishing security thus reduces to two steps: (i) certify a lower bound on $H_{\min}(\mathbf{X}|\mathbf{E})$, and (ii) apply privacy amplification to obtain the final random string.

\begin{table*}[ht!]
\renewcommand{\arraystretch}{1.5}
\setlength{\tabcolsep}{3pt}
\begin{tabular}{p{0.19\textwidth} p{0.27\textwidth} p{0.37\textwidth} >{\raggedleft\arraybackslash}p{0.12\textwidth}}
\toprule
\textbf{Protocol} & \textbf{Platform / assumption} & \textbf{Security core} & \textbf{Certified rate} \\
\midrule
Lunghi \textit{et al.} (2015)~\cite{Lunghi2015}
& DV; qubit dimension bound
& Dimension-bounded SDI; i.i.d.; self-testing witness $\rightarrow H_{\min}$ bound
& $\sim 23$ bits/s \\

Brask \textit{et al.} (2017)~\cite{Brask2017}
& Non-orthogonal coherent states; overlap bound; USD measurement
& Overlap-based SDI; i.i.d.
& $\sim 16.5$ Mbit/s \\

Marangon \textit{et al.} (2017)~\cite{marangon2017}
& CV homodyne; trusted detector, untrusted source
& Source-DI; entropic-uncertainty bound (random quadrature switching), i.i.d.
& $\sim 1.7$ Gbit/s \\

Rusca \textit{et al.} (2019)~\cite{Rusca2019}
& Weak coherent pulses + SPD; mean-energy bound (power meter)
& Vacuum component SDI; no i.i.d.; SDP bound on guessing probability
& $1.25$ Mbit/s \\

Avesani \textit{et al.} (2020)~\cite{Avesani2020}
& CV heterodyne (all-fibre); mean-energy bound
& Vacuum component SDI; no i.i.d.; SDP-based entropy bound
& $>113$ Mbit/s \\

Wang \textit{et al.} (2023)~\cite{Wang2023} 
& CV homodyne; trusted source, untrusted detector 
& Measurement-DI; no i.i.d. 
& $\sim 5.0$ Kbits/s \\

Bertapelle \textit{et al.} (2025)~\cite{Bertapelle2025} 
& CV heterodyne; trusted detector, untrusted source 
& Source-DI; entropic-uncertainty bound (random quadrature switching), i.i.d.
& $\sim 20$ Gbit/s

\end{tabular}

\caption{Overview of SDI-QRNG experiments. The table includes the  chosen encoding and decoding, the specific SDI-assumption and the security framework. The performance is estimated by the final certified extraction rate.}
\label{tab:sdiqrng}
\end{table*}

While in device-dependent-QRNG the entropy can be estimated directly from a detailed physical model of the devices, in SDI-QRNG the devices are only partially characterised. The entropy must therefore be bounded from observed correlations $p(b|x,y)$ together with the setup assumptions, in particular the bound on the communication (e.g. an energy bound). This typically leads to an optimisation problem of the general form
\begin{equation}
    \min H_{min}(\boldsymbol{X}|\boldsymbol{E}) \hspace{0.3cm}\text{s.t.}
\begin{cases}    
    \hspace{0.3cm} & \text{setup assumptions} \\
    & \text{reproducing }p(b|x,y)
\end{cases} .       
\end{equation}
Here, the optimisation is over all possible quantum realisations compatible with the constraints.

If the behaviour of the devices is independent and identically distributed (i.i.d.) across rounds, then it is sufficient to bound the single-round min-entropy. This simplifies computations, which in many cases can be solved efficiently with semidefinite programming \cite{PaulDani2023}. However, i.i.d.\ is a strong assumption, which is not easy to justify in practice. When the i.i.d.\ assumption is not desirable, different techniques are needed. In particular, one can bound $H_{\min}(\boldsymbol{X}|\boldsymbol{E})$ in terms of single-round entropies using entropy accumulation. First developed for DI-QKD \cite{Renner2009}, entropy-accumulation techniques have recently been generalised to prepare-and-measure scenarios \cite{Metger2023}. They also provide systematic methods to handle finite-size effects \cite{Roch2025,SDPreview}, i.e.~the fact that the correlations $p(b|x,y)$ are estimated from a finite data set.

Finally, it is generally desirable to establish security in a composable manner, meaning that the output random string is (up to a parameter $\epsilon$) indistinguishable from uniform and independent of the adversary even when used as a subroutine in a larger cryptographic task \cite{Renner2009,Tomamichel2016}. Operationally, composability is typically obtained by combining a bound on $H_{\min}(\boldsymbol{X}|\boldsymbol{E})$ with privacy amplification (e.g.\ via two-universal hashing or Toeplitz extraction) to produce an $\epsilon$-secure output string \cite{Renner2004,Tomamichel2011}.

Many different SDI models, based on different assumptions have been explored for QRNG, both theoretically and experimentally; Table~\ref{tab:sdiqrng} provides a selective overview. The first analysis has focused on the dimension-bounded model in the RAC scenario \cite{Li2011}, providing a lower bound on the min-entropy in terms of the RAC score. While conceptually clean, this approach is demanding in practice. First, as discussed in Sec.~III, directly verifiable in an adversarial setting—especially in optics, where extra degrees of freedom (multi-photon components, spectral/temporal modes, spatial structure) are hard to exclude and can inadvertently leak information. Second, the certification of randomness in the dimension-bounded model typically requires high detection efficiency \cite{dallarno2015,Sarubi2025,Sanz2026}, comparable to fully DI protocols. In principle, this can be addressed by placing an additional assumption of ``fair sampling’’, but this is difficult to motivate in an adversarial SDI scenario. This motivates exploring alternative models. 

A first step towards a practical implementation was reported in Ref.~\cite{Lunghi2015}, with an optical QRNG experiment based on polarization photons, assuming independent preparation and measurement devices. This protocol is robust to loss, and randomness generation rates of 23 bits/s could be reported with standard optical components and detectors. An interesting aspect of such a setup is its ``self-testing’’ ability. This means that the user can verify in real-time the generation of certified random bits, via the continuous entropy estimation based on the observed correlations. 

In turn significantly improved QRNG schemes have been proposed by introducing the energy and overlap SDI models, hence moving away from the dimension-bounded model. The improvement is two-fold. First, the assumption underlying security is much better tailored to experiments, and verifiable in practice. Second, the scheme is simpler, as it can be implemented via a single (fixed) measurement. Schemes based on standard optical setups with weak coherent states and homodyne measurements. This leads to increased performance with rates in the Mbit/s regime. 

While these schemes require a partial characterization of the preparation device, another approach has been developed where randomness can be certified without placing any assumption on the source. In this source-DI model, it is the measurement device that is assumed to be fully characterized. Here randomness can be quantified via entropic uncertainty relations \cite{Vallone2014}, and implementations based on homodyne and heterodyne measurements have been reported, see for example \cite{Cao2016,Avesani2018,Pivoluska2021,Cheng2024}. The converse scenario, where the source is trusted but the measurement device is untrusted (referred to as measurement-DI) has also been proposed \cite{Cao2015} and implemented \cite{Nie2016,Wang2023}.

The prepare-and-measure setup also underlies quantum key distribution (QKD), where two honest parties, Alice and Bob, aim to establish a shared secret key that is unpredictable to any eavesdropper. The paradigmatic protocol is BB84 \cite{Bennet1984}: Alice encodes bits in two mutually unbiased bases (typically Pauli $X$ and $Z$), Bob measures in a randomly chosen basis, and they later announce basis choices publicly to sift the raw key. Parameter estimation reveals the disturbance level (hence potential eavesdropping), after which error correction and privacy amplification are used to produce a final secret key. While BB84 is secure in its idealised model, practical implementations require additional trust and calibration assumptions, which has motivated extensive work on implementation security and side-channel attacks, notably on the detection stage (e.g.\ detector blinding)~\cite{Lydersen2010}.

This naturally raises the same question as for QRNG: which parts of the implementation must be trusted for security, and what can instead be certified from the observed statistics under natural, experimentally verifiable assumptions? A first route is to impose a bound on the Hilbert-space dimension of the communicated signals. For BB84 as usually implemented, this is not sufficient to certify a quantum advantage at the level of the observed prepare-and-measure correlations, since these correlations can be simulated with a classical bit. However, by rotating the measurement bases (e.g.\ to those used in the random access code task of Sec.~II), one obtains correlations that do witness quantumness and can be used for semi-device-independent security~\cite{Pawlowski2011}. Moreover, as we discussed in Sec.~II, the RAC can be viewed as a prepare-and-measure analogue of CHSH Bell inequality, and a closely related entropy--violation trade-off carries over~\cite{Woodhead2015}. Motivated by this analogy, one can lower bound the conditional min-entropy $H_{\min}(A|E)$ as a function of the observed RAC score $P_{RAC}$, 
\begin{equation}
H_{\min}(A|E)\;\ge\;1-\log_2\!\Bigl(1+\sqrt{2-S^2/4}\Bigr),
\end{equation}
with $S=4(2P_{RAC}-1)$. The formula is the same for the case of DIQKD, where $S$ denotes the CHSH Bell value; note however that the derivation relies on a different set of tools.

Similarly to the case of QRNG, it is also desirable to go beyond the case of the dimension bounded model. In particular, since QKD typically involves large distances (and hence high loss) the requirements in terms of detection efficiency become even more important. This motivates exploring alternative SDI assumptions that are better aligned with realistic photonic implementations. A recent proposal is receiver-device-independent QKD~\cite{Ioannou2022}: Alice’s source is partially characterised through bounds on pairwise state overlaps (equivalently, constraints on the Gram matrix of the prepared states), while Bob’s measurement device is treated as an untrusted black box. This framework is immune to detector-side attacks such as blinding, avoids an explicit dimension assumption, and can yield positive key rates even at very low channel transmission when Alice prepares sufficiently many distinct states. A proof-of-principle experiment using weak coherent states and standard telecom components demonstrated secure key distribution over a 5km  fibre~\cite{Ioannou2022exp}. Interestingly, while SDI prepare-and-measure QKD was historically inspired by entanglement-based (and ultimately device-independent) protocols, prepare-and-measure results can also be used in the opposite direction as building blocks towards fully device-independent QKD~\cite{Sekatski2025}.

It is interesting to investigate other possible applications in the SDI model, beyond QRNG and QKD. First steps have been taken, considering SDI schemes for quantum money \cite{bozzio2019,Horodecki2020} and quantum private queries \cite{Basak2023QIP}.

\section{Conclusion and outlook}

Prepare-and-measure scenarios provide a minimal setting in which to investigate quantum systems as information carriers. In this review we covered three main themes: quantum-over-classical advantages under various communication constraints, the geometry of the corresponding classical and quantum correlation sets, and SDI applications such as certification, randomness generation, and QKD.

The central question underlying this research is that of a quantum advantage. This has been most thoroughly investigated in the dimension-bounded model. Yet some fundamental questions remain open, in particular whether the classical simulation cost of high-dimensional quantum systems is finite or not. Quantum advantages in other SDI models remain largely unexplored — an interesting avenue for future research, particularly in finding instances of large quantum-to-classical gaps featuring strong robustness to noise and loss. Beyond their conceptual interest, such instances could provide the basis for novel SDI protocols, especially for QKD.

Looking ahead, we expect the insights gained from prepare-and-measure scenarios to prove valuable as these ideas are being extended to entanglement-assisted \cite{Tavakoli2021, Pauwels2022}, sequential \cite{Chen2024}, time-evolution \cite{Konstantinos2025}, instrumental \cite{Chaves2018,VanHimbeeck2019} and network \cite{Bowles2015, Lauand2025} scenarios. We hope this review will motivate further developments on these exciting topics.

\acknowledgements
A.T.~is supported by the Knut and Alice Wallenberg Foundation through the Wallenberg Center for Quantum Technology (WACQT), the Swedish Research Council under Contract No.~2023-03498 and  the Swedish Foundation for Strategic Research. N.B~and J.P~acknowledge support from the Swiss
State Secretariat for Education, Research and Innovation (SERI) under contract number UeM019-3.
D.R.~acknowledges support from the Galician Regional Government (consolidation of Research Units: AtlantTIC and own funding through the ``Planes Complementarios de I+D+I con las Comunidades Autonomas'' in Quantum Communication), MICIN with funding from the European Union NextGenerationEU (PRTR-C17.I1), the ``Hub Nacional de Excelencia en Comunicaciones Cu{\' a}nticas'' funded by the Spanish Ministry for Digital Transformation and the Public Service and the European Union NextGenerationEU, the European Union's Horizon Europe Framework Programme under Marie Sk\l{}odowska-Curie grant 101072637 (project QSI) and project ``Quantum Security Networks Partnership'' (QSNP; grant 101114043), and the European Union via the European Health and Digital Executive Agency (HADEA) under project QuTechSpace (grant 101135225). J.B.B.~acknowledges support from the Danish National Research Foundation
grant bigQ (DNRF 142) and the Innovation Fund Denmark Grand Solutions grant 3200-00012B TripleQ.

\bibliographystyle{apsrev4-2}
\bibliography{references}	

\end{document}